\documentclass[showpacs,preprintnumbers,amsmath,amssymb,superscriptaddress]{revtex4}

\usepackage{graphicx}
\usepackage{dcolumn}
\usepackage{bm}
\usepackage{color}

\begin{document}

\preprint{Phys. Rev. E \textbf{79}, 062901 (2009).}

\title{Four-state rock-paper-scissors games in constrained Newman-Watts networks}

\author{Guo-Yong Zhang}
\affiliation{Institute of Theoretical Physics, Lanzhou University, Lanzhou 730000, China}
\affiliation{Department of Computer Science, Hubei Normal University, Huangshi 435002, China}

\author{Yong Chen}
\altaffiliation{Corresponding author. Email: \tt{ychen@lzu.edu.cn}}
\affiliation{Institute of Theoretical Physics, Lanzhou University, Lanzhou 730000, China}

\author{Wei-Kai Qi}
\affiliation{Institute of Theoretical Physics, Lanzhou University, Lanzhou 730000, China}
\affiliation{Department of Industrial Systems and Engineering, The Hong Kong polytechnic University, Hung Hom, Kowloon, Hong Kong, China }

\author{Shao-Meng Qing}
\affiliation{Institute of Theoretical Physics, Lanzhou University, Lanzhou 730000, China}

\date{\today}

\begin{abstract}
We study the cyclic dominance of three species in two-dimensional constrained Newman-Watts networks with a four-state variant of the rock-paper-scissors game. By limiting the maximal connection distance $R_{max}$ in Newman-Watts networks with the long-rang connection probability $p$, we depict more realistically the stochastic interactions among species within ecosystems. When we fix mobility and vary the value of $p$ or $R_{max}$, the Monte Carlo simulations show that the spiral waves grow in size, and the system becomes unstable and biodiversity is lost with increasing $p$ or $R_{max}$. These results are similar to recent results of Reichenbach \textit{et al.} [Nature (London) \textbf{448}, 1046 (2007)], in which they increase the mobility only without including long-range interactions. We compared extinctions with or without long-range connections and computed spatial correlation functions and correlation length. We conclude that  long-range connections could improve the mobility of species, drastically changing their crossover to extinction and making the system more unstable.
\end{abstract}

\pacs{87.23.Cc, 89.75.Fb, 05.50.+q}
\maketitle

The question of how biological diversity is maintained has initiated increasingly more research from multiple angles in recent decades~\cite{frean,kerr,reichenbach,mobilia,claussen,murray}. Mathematical modeling of population dynamics is widely recognized as a useful tool in the study of many interesting features of ecological systems. However, the enormous number of interacting species found in the Earth's ecosystems is a major challenge for theoretical description. For this reason, researchers have built many simplified models to describe the evolution of ecological systems over time~\cite{lotka,volterra,kerr,durrett,windus,hastings}. One of the simplest cases is of three species that have relationships analogous to the game of rock-paper-scissors (RPS), where rock smashes scissors, scissors cut paper, and paper wraps rock. It is a well-studied model of population dynamics~\cite{matti,szabo1,szolnoki,szabo2,efimov}, and it can be classified in two ways: three-state or four-state models, depending on whether we consider the empty state or not. It is well known that such a cyclic dominance can lead to nontrivial spatial patterns as well as coexistence of all three species.

Recently, Reichenbach and co-workers studied a stochastic spatial variant of the RPS game~\cite{reichenbach,mobilia,tobias}. In their study, they run the game with four states: the three original cyclically dominating states and a fourth one that denotes empty space. In addition, they introduced a form of mobility to mimic a central feature of real ecosystems: animal migration, bacteria run and tumble. They found that mobility has a critical influence on species diversity~\cite{reichenbach,mobilia,tobias}. When mobility exceeds a certain value, biodiversity is jeopardized and lost. In contrast, below this critical threshold value, spatial patterns can form and help  enable stable species diversity. We shall take this population model as a basis to construct a new version of the three-species food chain in the constrained Newman-Watts (NW) networks. In the model studied by Reichenbach and co-workers, they consider mobile individuals of three species (referred to as $A$, $B$, and $C$), arranged on a spatial lattice, where each individual can only interact with its nearest neighbors. In this study, we introduce some stochastic long-range interactions between elements of the lattice. The stochastic long-range interactions occur when there exist long-range connections in the NW networks, mimicking a more real ecosystem: e.g., birds can fly, so they can prey not only near their nest but also at longer distances from the nest~\cite{sabrina}, pathogens disperse by air and water~\cite{brown,mccallum}, biological invasions related to human influence occur over long distances~\cite{ruiz}, and the long-range dispersal of plant seeds is driven by large and migratory animals, ocean currents and human transportation~\cite{nathan}. Of course, the long-range interaction cannot be infinite, so we limit the distance of long-range interactions to $R_{max}$. That is, the individuals are assigned an interaction distance. For the sake of simplicity, we consider that the maximum interaction distance $R_{max}$ and the long-range interaction probability $p$ are the same for all species. With Monte Carlo (MC) simulations we show that the maximum interaction distance $R_{max}$ and the long-range interaction probability $p$ play an important role in the coexistence of all three species.

We consider the four-state RPS model which was described in detail in Refs.~\cite{reichenbach,tobias,matti}. Here, we give a recapitulation:
\begin{eqnarray}
AB\stackrel{\sigma}\rightarrow AE, \quad & BC\stackrel{\sigma}\rightarrow BE, \quad &  CA\stackrel{\sigma}\rightarrow CE. \nonumber \\
AE\stackrel{\mu}\rightarrow AA, \quad & BE\stackrel{\mu}\rightarrow BB, \quad &  CE\stackrel{\mu}\rightarrow CC.
\label{eq01}
\end{eqnarray}
Here, $A$, $B$, and $C$ denote the three species which cyclically dominate each other, and $E$ denotes an available empty space. An individual of species $A$ can \textit{kill} B, with successive production of E. Cyclic dominance occurs as A can \textit{kill} B, B preys on C, and C beats A in turn, closing the cycle. These processes are called `selection' and occur at a rate $\sigma$. To mimic a finite carrying capacity, each species can reproduce only if an empty space is available, as described by the reaction $AE \rightarrow AA$ and analogously for $B$ and $C$. For all species, these reproduction events occur at a rate $\mu$.

In addition, to mimic the possibility of migration, one can amend the reaction equations with an exchange reaction:
\begin{eqnarray}
XY\stackrel{\epsilon}\rightarrow YX.
\label{eq02}
\end{eqnarray}
where $X$ and $Y$ denote any state (including empty space) and $\epsilon$ is the exchange rate. The mobility was defined as $m=2 \epsilon N^{-1}$ in Ref.~\cite{reichenbach}, where $N$ denotes the number of sites. From a dynamic viewpoint, the RPS game can be described by the mean field rate equations~\cite{tobias,matti},
\begin{eqnarray}
\partial_{t} a &=& a[\mu(1-\rho)-\sigma c], \nonumber \\
\partial_{t} b &=& a[\mu(1-\rho)-\sigma a], \nonumber \\
\partial_{t} c &=& a[\mu(1-\rho)-\sigma b],
\label{eq03}
\end{eqnarray}
where $a$, $b$, and $c$ are densities of the states $A$, $B$, and $C$, respectively. That is,
\begin{equation}
a=N_a/N, \quad  b=N_b/N, \quad c=N_c/N,
\label{eq04}
\end{equation}
where $N_a$, $N_b$, and $N_c$ are the number of species of $A$, $B$, and $C$, respectively. $\rho=a+b+c$ is the total density. These equations have a reactive fixed point $a=b=c=\frac{\mu}{3\mu+\sigma}$, which is linearly unstable~\cite{tobias}.

The mean field approach does not take into account the spatial structure and assumes the system to be well mixed. Therefore, it can only serve as a rough model for dynamic processes. Here, we consider the spatial version of the above model in the complex NW network structure~\cite{newman}, and we use the Monte Carlo simulation approach. The two-dimension (2D) NW network was constructed as follows: (i) We first built a 2D $L \times L$ $(N=L^2)$ regular square lattice. So, the total number of connections is $2N$. (ii) Then, we randomly chose two sites that have no direct connection. If the shortest path length between the two sites was shorter than the maximal distance $R_{max}$, we connected the sites; if not, we choose other two sites, until the number of the long-range connections equaled $2pN$. Here, the shortest path length refers to that we did not take into account the long-range connections. 

\begin{figure}
\includegraphics[width=0.4\textwidth]{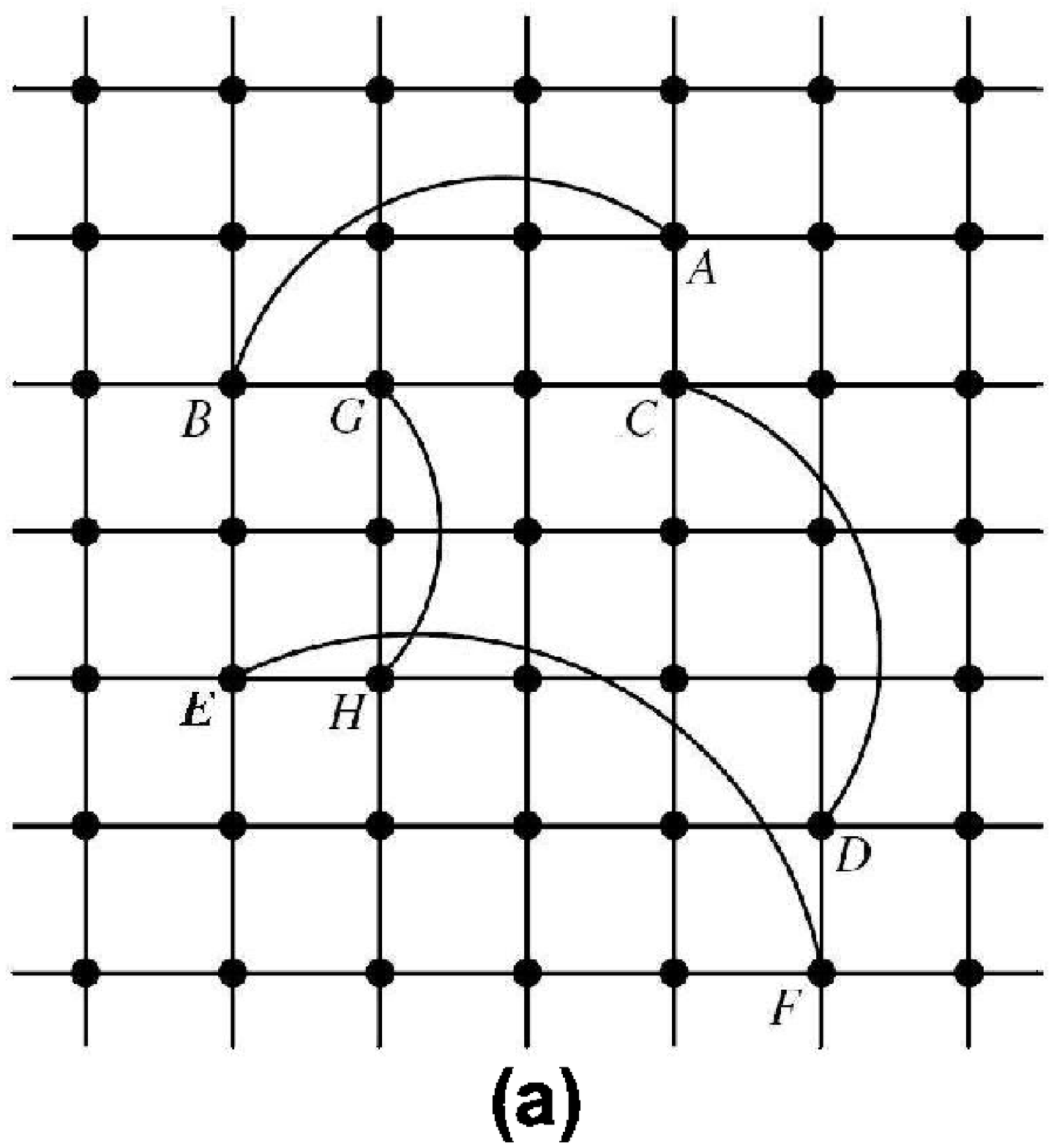}
\includegraphics[width=0.4\textwidth]{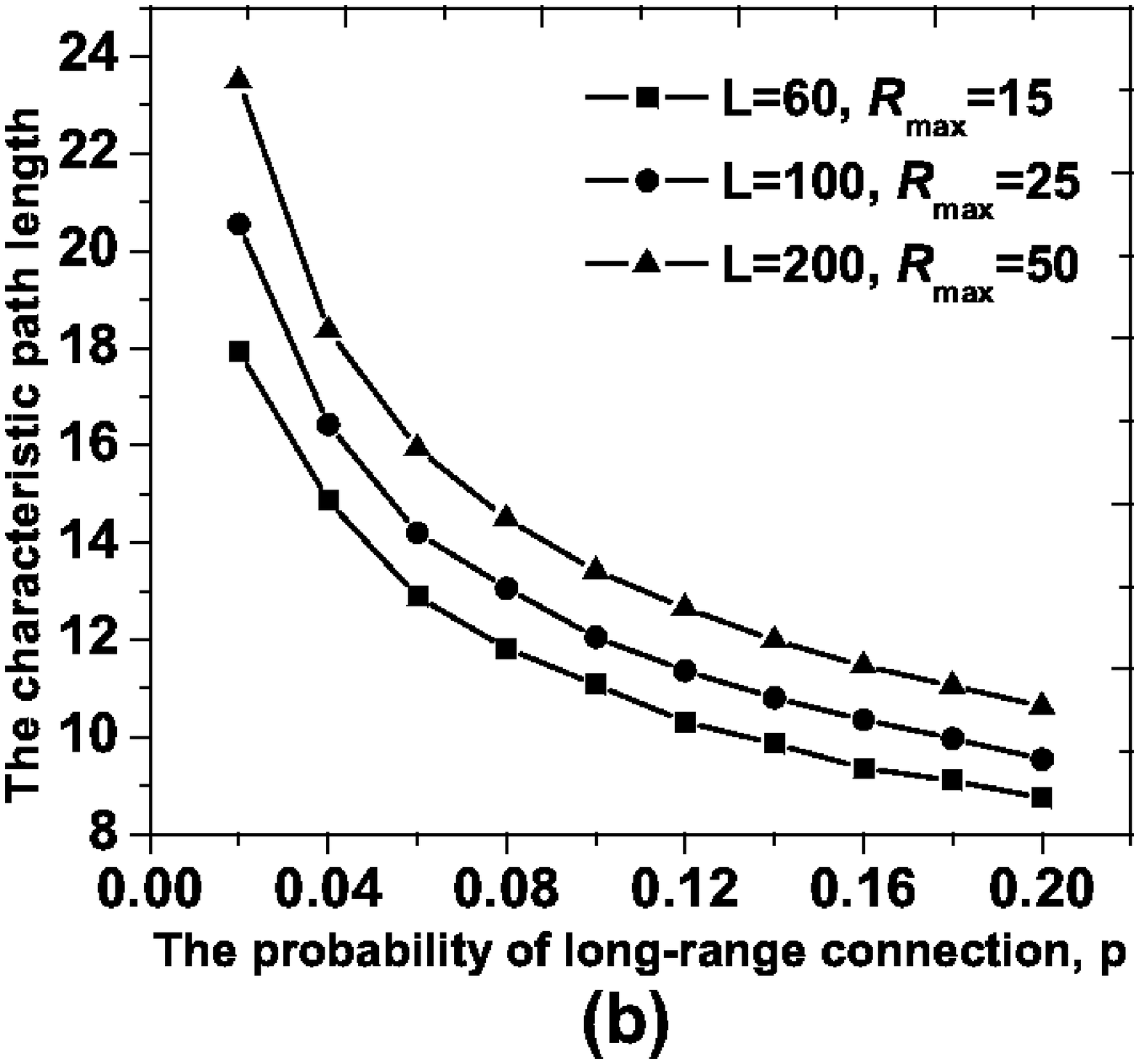}
\caption{(a) The structure of a constrained NW network. All the long-range connections within the range $[1,R_{max}]$ and the range of $R_{max}$ is $1 \leq R_{max} \leq L$. (b) The characteristic path length as a function of the long-range connection probability $p$. The maximal long-range connection distance $R_{max} = L/4$.}
\label{fig01}
\end{figure}

The above procedure produces a constrained NW network structure as shown in Fig.~\ref{fig01}(a). In Fig.~\ref{fig01}(b), we plot the characteristic path length as a function of the long-range connection probability $p$. The characteristic path length decreases with an increase in the long-range connection probability $p$. The long-range connection probability $p$ is equivalent to the rewiring probability in a Watts-Strogatz network but connections are added without removing any of the original ones. So, the modified NW structure is characterized by the probability $p$ and the maximal long-range connection distance $R_{max}$. Once the network was built as described above, the evolution of the system over time obeyed the following rules (modified from Refs.~\cite{reichenbach,tobias}): (i) Consider mobile individuals of three species (referred to as $A$, $B$, and $C$), scattered randomly on a square lattice as in Fig.~\ref{fig01}(a) with periodic boundary conditions. Every lattice site was initially occupied by an individual of species $A$, $B$, or $C$, or left empty. (ii) At each simulation step, a random individual was chosen to interact with a randomly-chosen individual directly connected to it. A process (selection, reproduction, or mobility) was chosen randomly with a probability proportional to the rates, and the corresponding reaction is executed.

In the above process, $N=L^2$ simulation steps constitute one Monte Carlo step (MCS). During one MCS all lattice sites had one chance to interact. Over time, the spatial distributions of $A$, $B$, and $C$ species changed from one MCS to another, providing the evolution of the system at the microscopic level. 

According to Ref.~\cite{tobias}, Eq.~(\ref{eq03}) could be cast into the form of the complex Ginzburg-Landau equation (CGLE). In accordance with the known behaviors of the CGLE, it was found that the spatial four-state model with diffusion leads to the formation of spirals. The spirals' wavelength $\lambda$ is proportional to the square root of mobility~\cite{reichenbach,tobias}. To investigate how the long-range interaction probability $p$ and the maximal distance $R_{max}$ affect the behaviors of the four-state model, we ran MC simulations of this model in the constrained NW network with periodic boundary conditions. All the results that we present were obtained starting from a random initial distribution of individuals and vacancies, and each site was in one of the four possible states. The densities of $A$, $B$, and $C$ coincided with the values of the unstable reactive fixed point of the rate equations~(\ref{eq03}). We considered equal selection and reproduction rates, which were set $\mu = \sigma = 1$~\cite{tobias}. So, all four states initially occurred with equal probability $1/4$.

\begin{figure}
\includegraphics[width=0.2\textwidth]{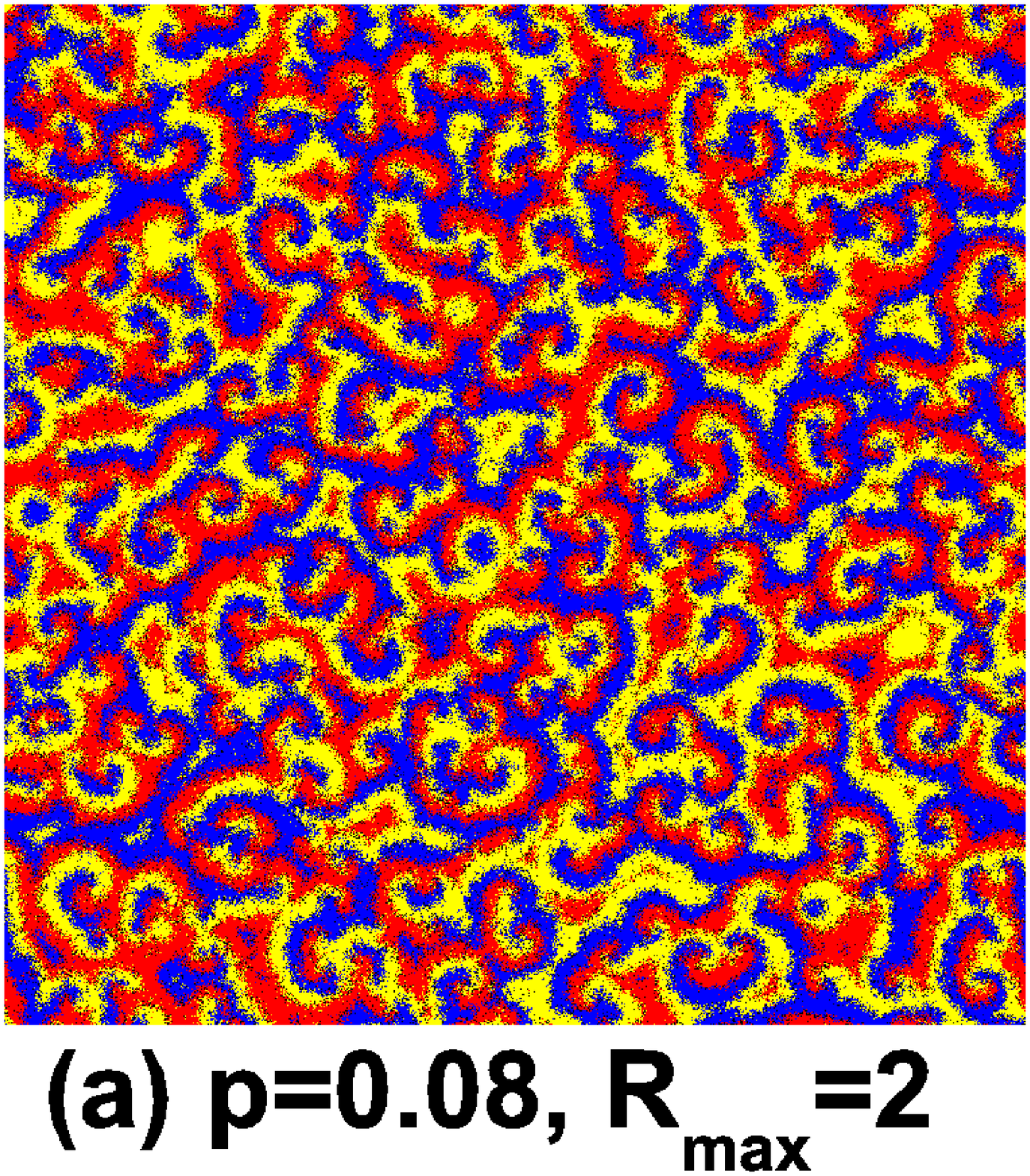}
\includegraphics[width=0.2\textwidth]{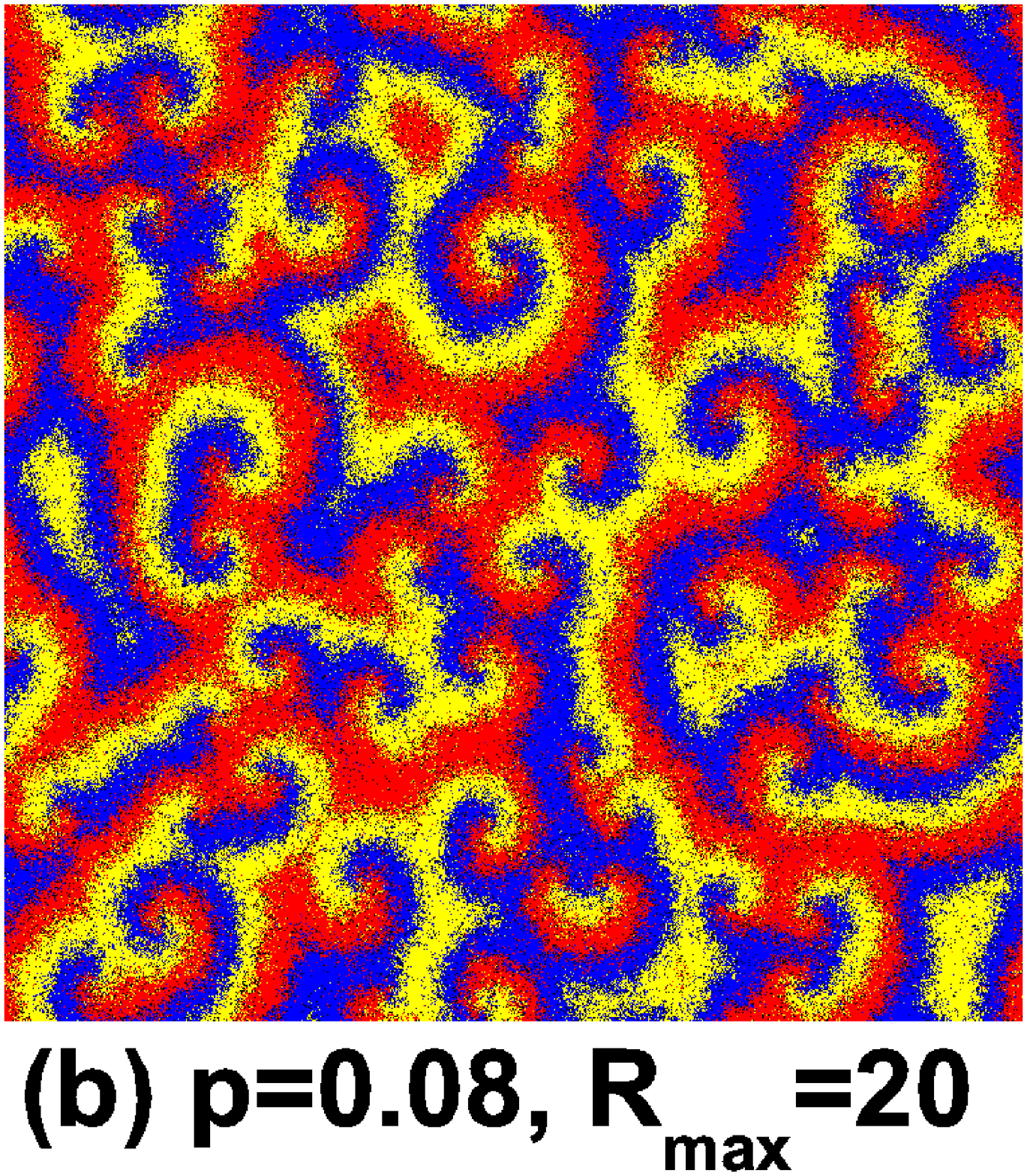}
\includegraphics[width=0.2\textwidth]{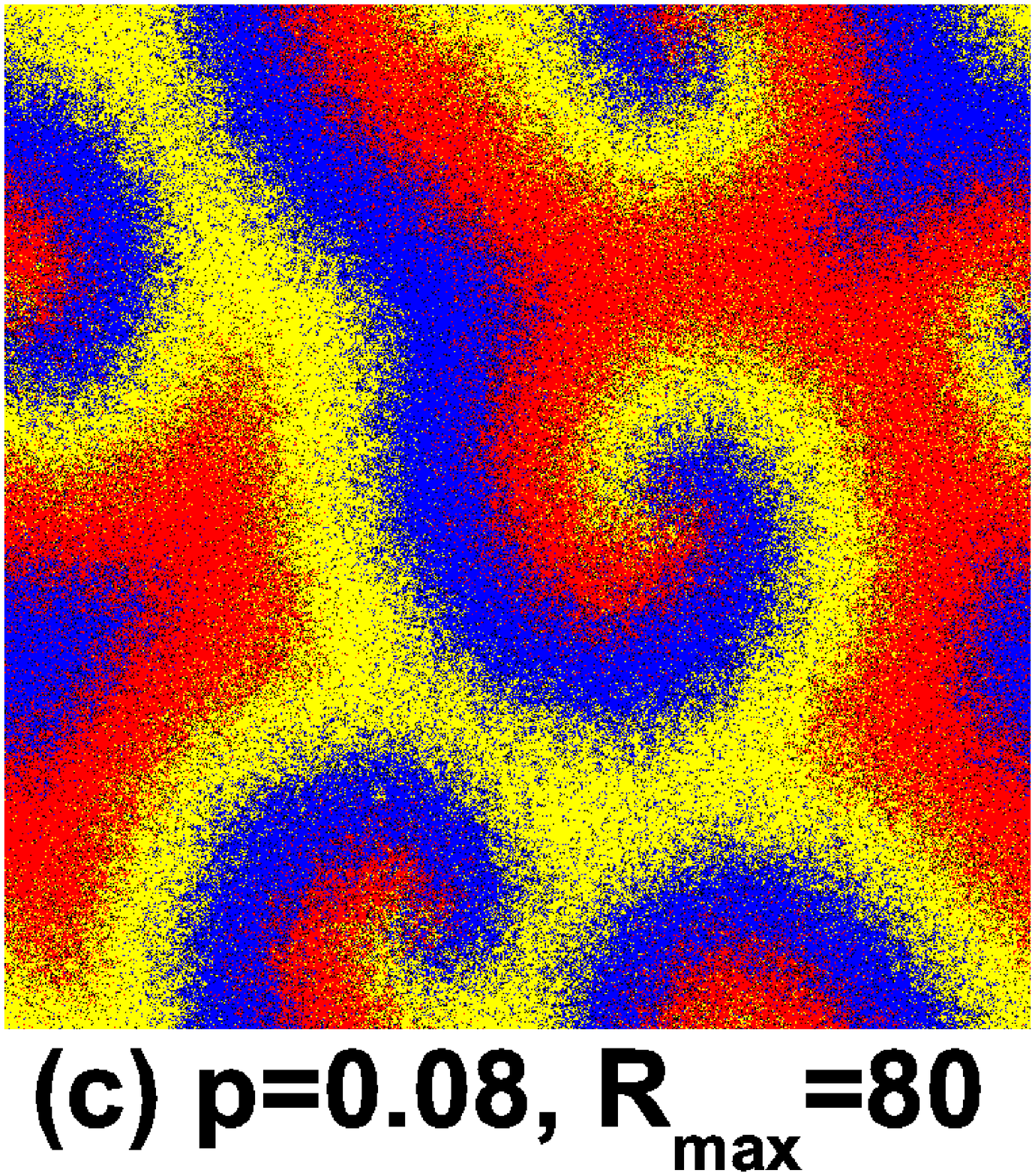}
\includegraphics[width=0.2\textwidth]{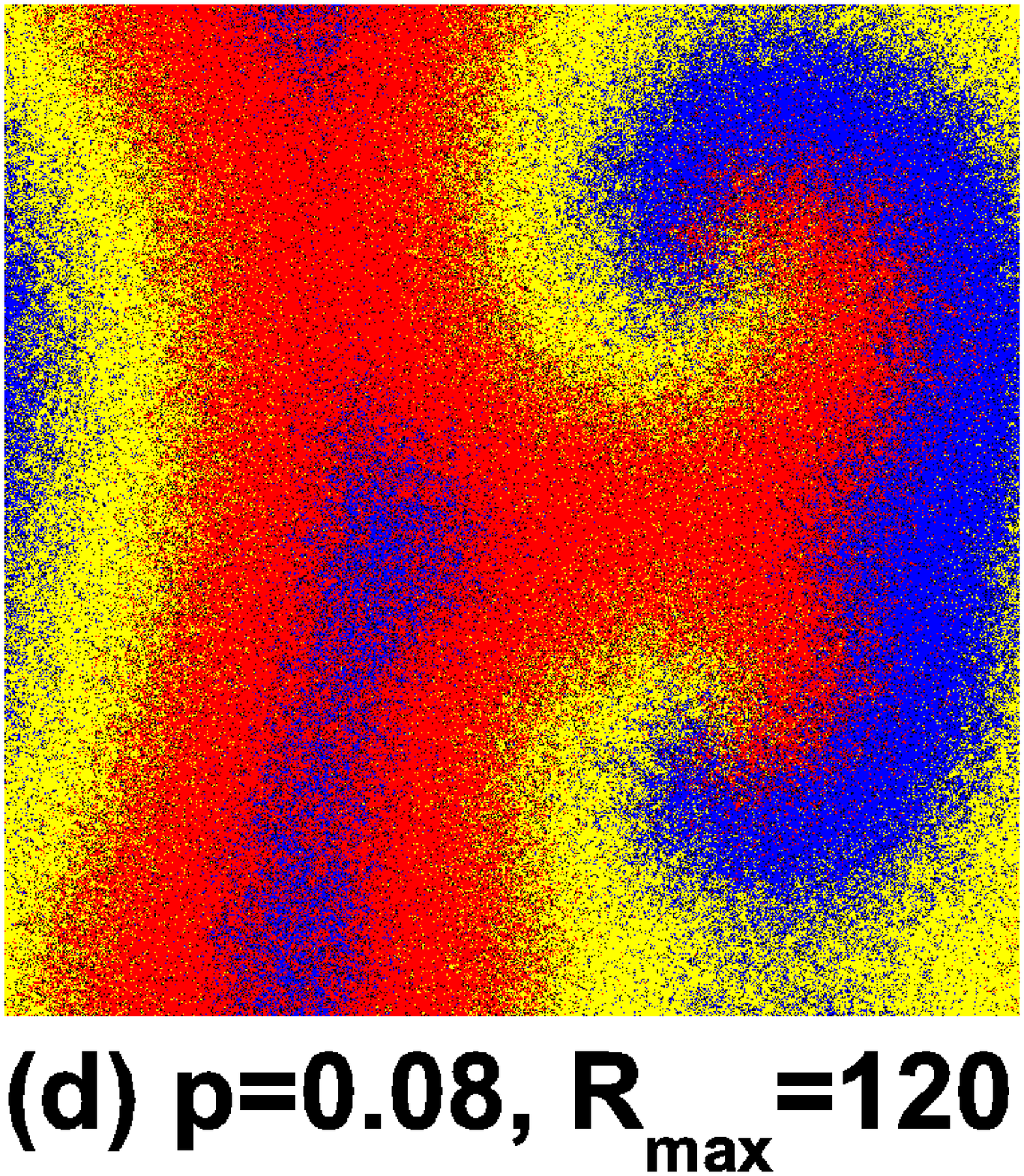}
\caption{(Color online) Snapshots of the reactive steady state for $m = 4 \times 10^{-6}$, $\mu = \sigma = 1$, and system size $L = 1000$ ($\epsilon = 2$). The long-range connection probability $p=0.08$, and the maximal interaction distance increases from $R_{max}=2$ to $120$.}
\label{fig02}
\end{figure}

\begin{figure}
\includegraphics[width=0.4\textwidth]{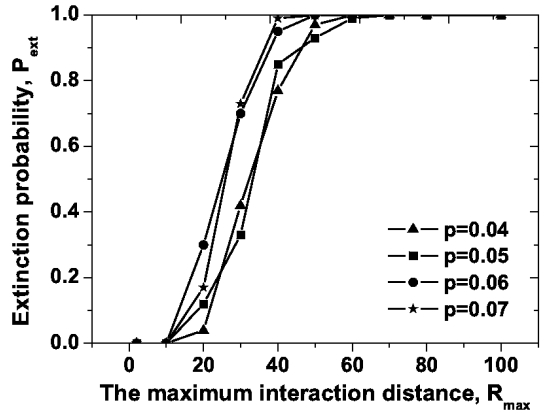}
\caption{Extinction probability as a function of the maximal long-range interaction distance $R_{max}$. Parameters: $L = 200$, $t = 10,000$ MCS, $\mu = \sigma = 1$, and $m = 1 \times 10^{-4}$. As $R_{max}$ increases, the transition from stable coexistence ($P_{ext} = 0$) to extinction ($P_{ext}=1$) sharpens at a critical value $R_c \approx 30$.}
\label{fig03}
\end{figure}

In Fig.~\ref{fig02}, we plot typical snapshots of the reactive steady states for various values of the maximal interaction distance $R_{max}$. When $R_{max}$ is short, long-range interactions have little effect, and all species coexist. The pattern of spiral waves in Fig.~\ref{fig02}(a) is similar to the case without long-range interactions. With increasing $R_{max}$, the spiral waves grow in size and eventually disappear for longer enough values of $R_{max}$. When the spiral waves disappear, the system becomes a uniform state where only one species exists and the others have died out. This process is similar to the result from increased mobility $m$ in the lattice simulation without long-range interactions in Ref.~\cite{reichenbach}. In addition, we computed the extinction probability $P_{ext}$ that two species have gone extinct after $10000$ MCS (see Fig.~\ref{fig03}). Fig.~\ref{fig03} clearly shows that there exists a critical value $R_c \approx 30$ in the process of phase transition from coexistence ($P_{ext}$ tends to zero) to extinction ($P_{ext}$ approaches $1$). Of course, the critical value $R_c$ depends on the other parameters, such as the system size, mobility, and so on.

\begin{figure}
\includegraphics[width=0.2\textwidth]{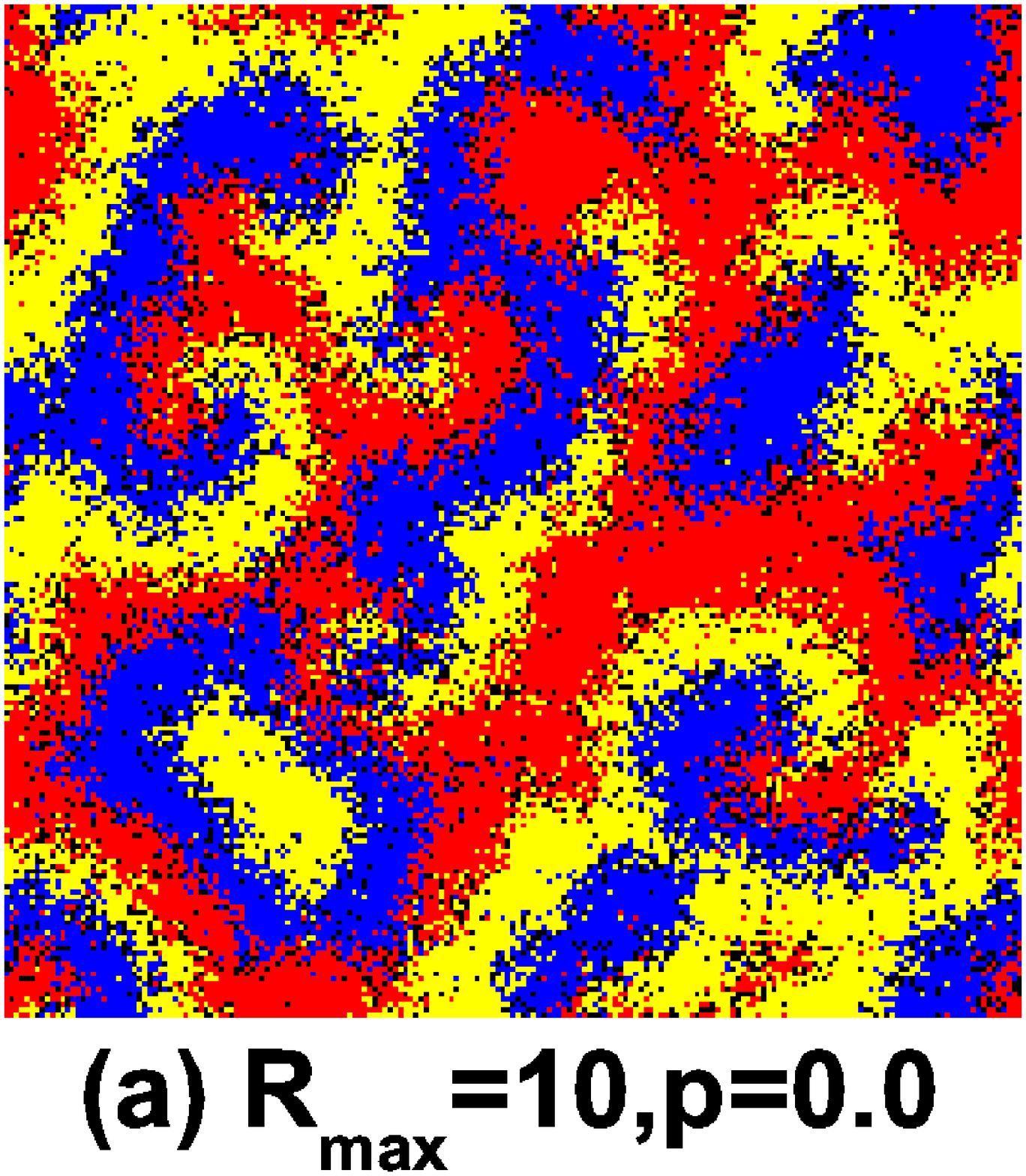}
\includegraphics[width=0.2\textwidth]{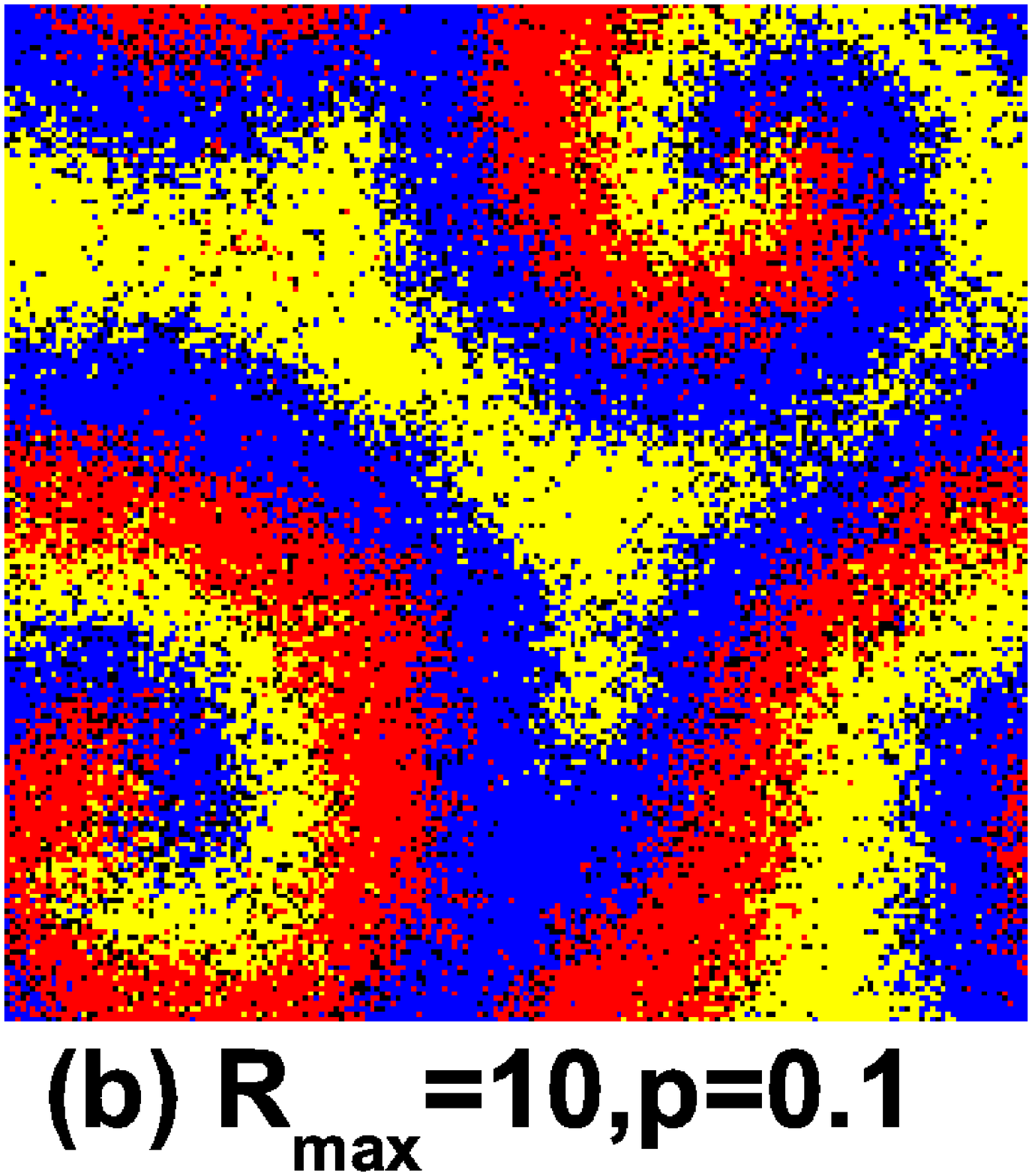}
\includegraphics[width=0.2\textwidth]{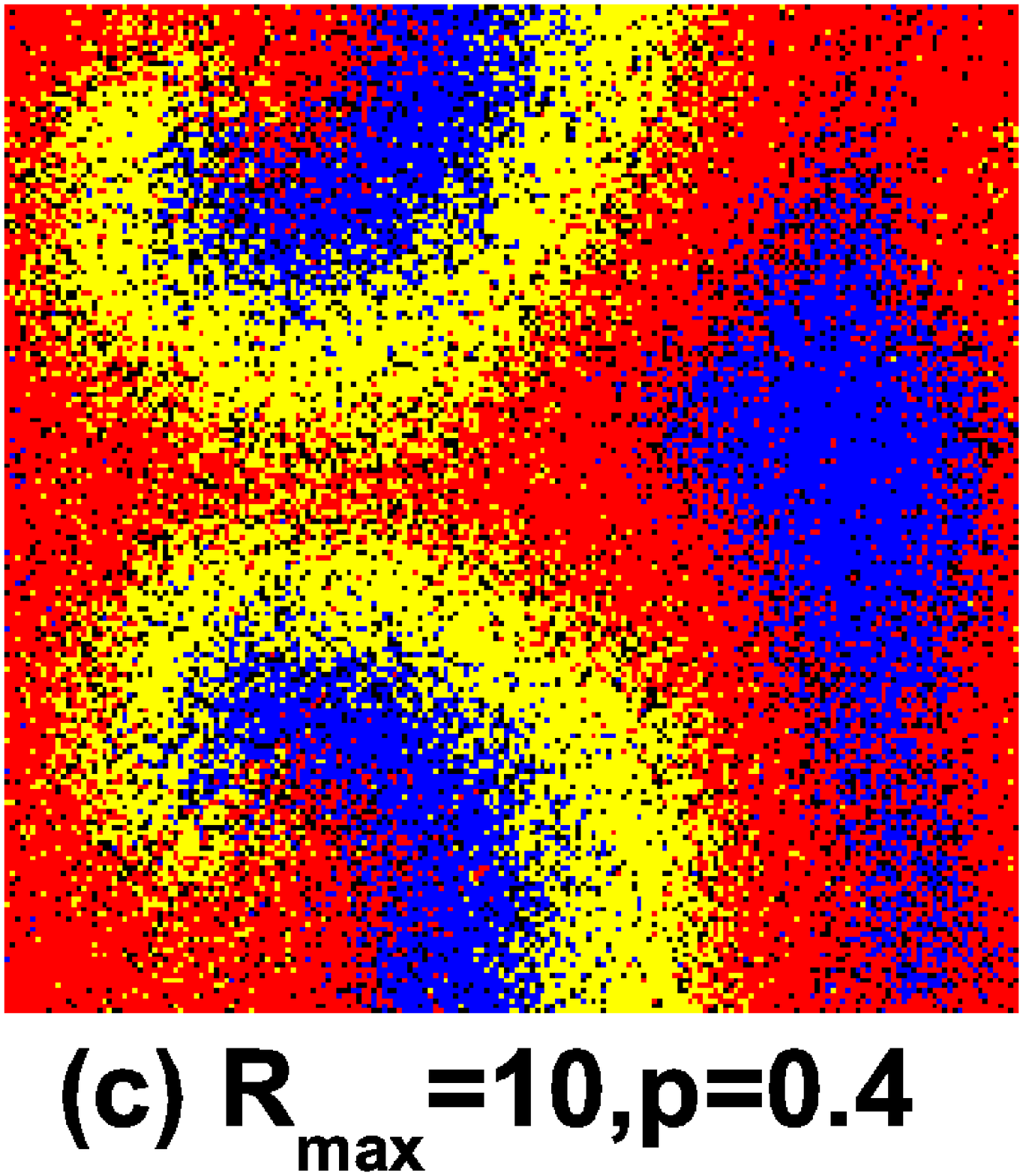}
\includegraphics[width=0.2\textwidth]{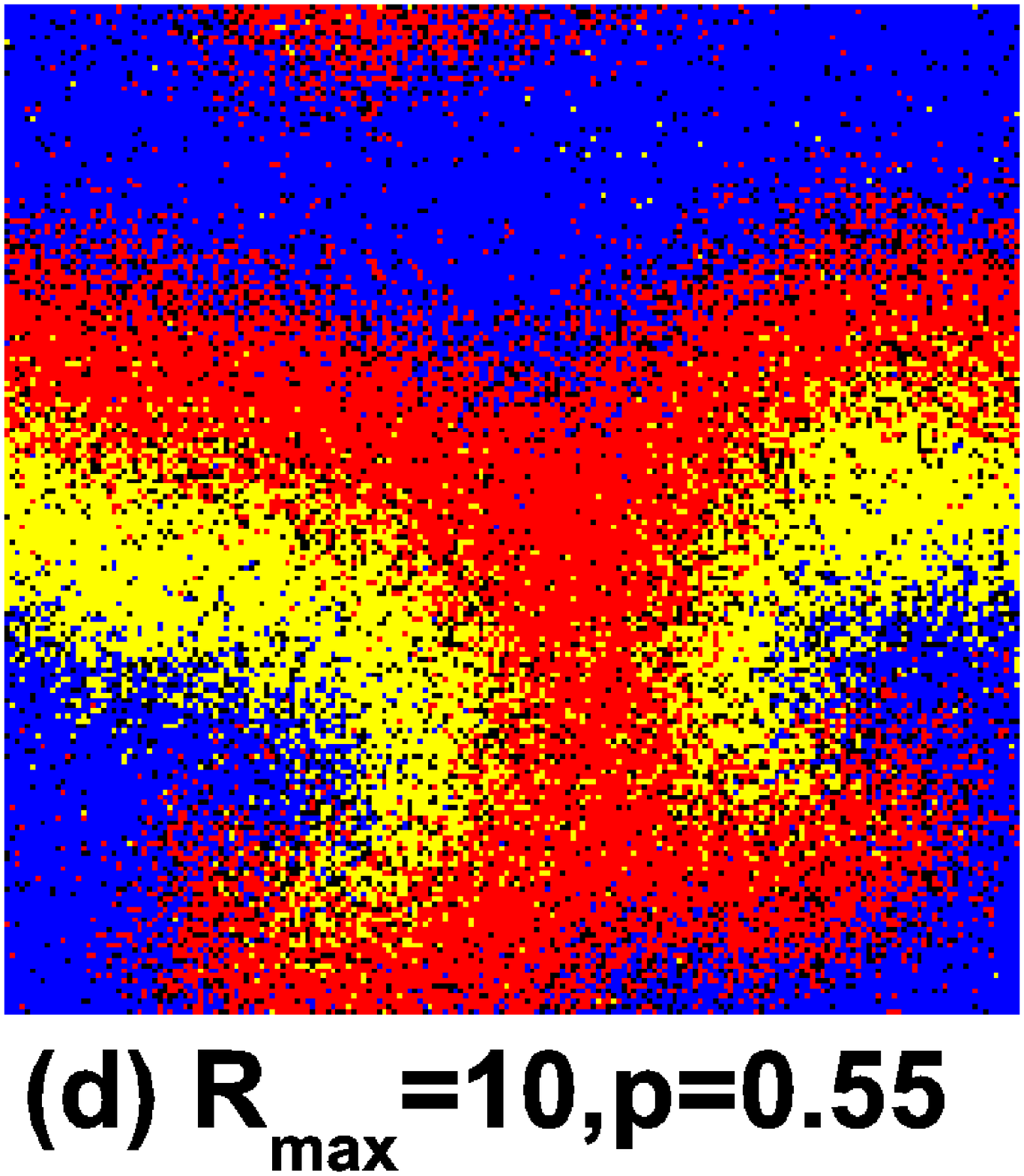}
\caption{(Color online) Snapshots of the reactive steady state for $m = 1 \times 10^{-4}$, $\mu = \sigma = 1$, and system size $L = 200$ ($\epsilon = 2$). The fixed maximal interaction distance $R_{max}=10$, and the long-range connection probability increases from $p=0.0$ to $0.55$.}
\label{fig04}
\end{figure}

\begin{figure}
\includegraphics[width=0.4\textwidth]{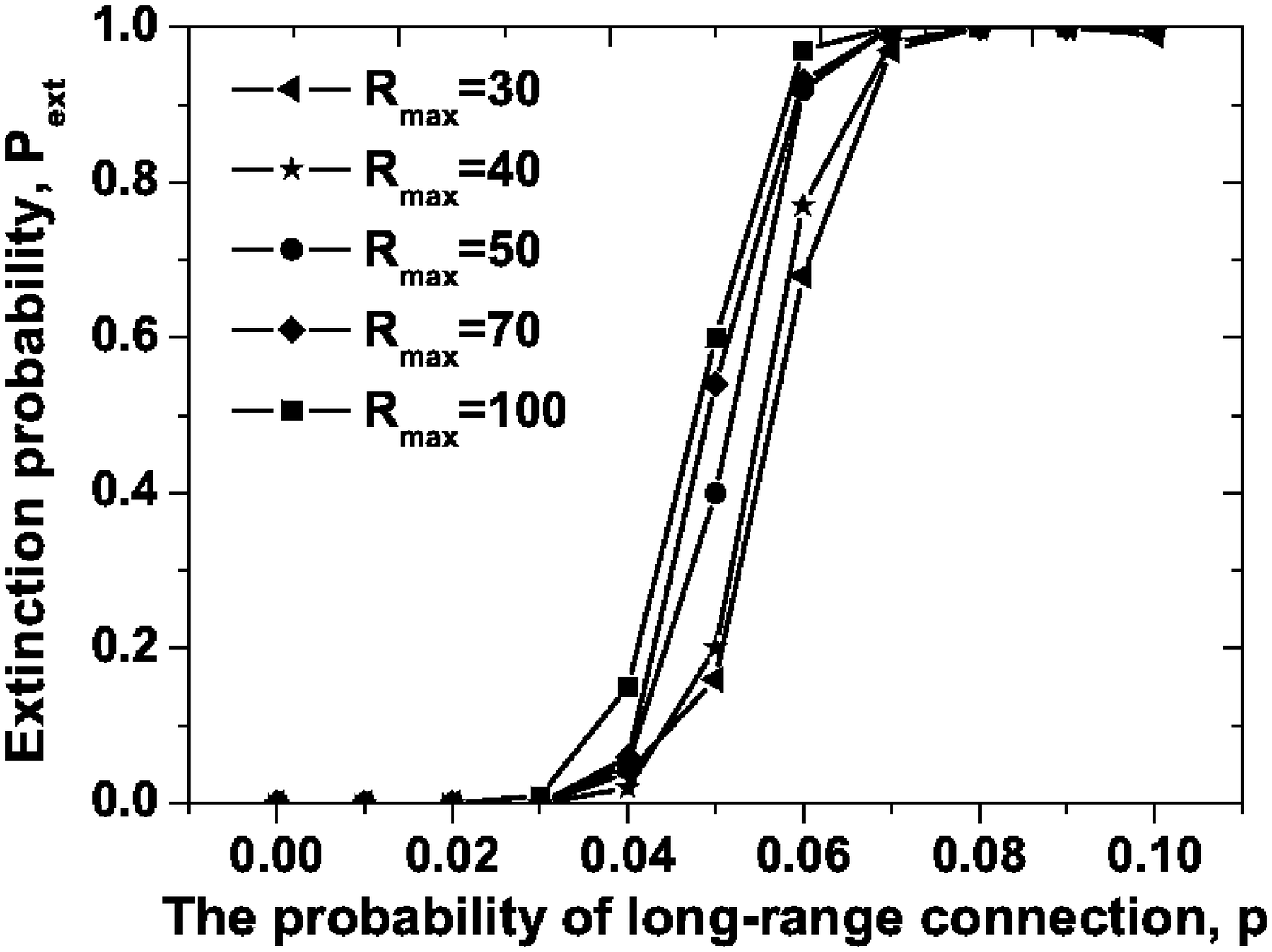}
\caption{The extinction probability as a function of the long-range connection probability $p$. Parameters: $L = 100$, $t = 10,000$ MCS, $\mu = \sigma = 1$, and $m = 2 \times 10^{-5}$. As $p$ increases, the transition from stable coexistence ($P_{ext} = 0$) to extinction ($P_{ext}=1$) sharpens at a critical value $p_c \approx 0.06$.}
\label{fig05}
\end{figure}

To investigate how the long-range connection probability $p$ affects the coexistence, we fixed the maximal long-range interaction distance to $R_{max}=10$ and varied the probability $p$. The simulation results are shown in Fig.~\ref{fig04}, and the dependence of the extinction probability $P_{ext}$ on $p$ is plotted in Fig.~\ref{fig05}. It turns out that $p$ has effects similar to those of $R_{max}$ on the extinction probability and spiral wave pattern. There also exists a critical value $p_c \approx 0.06$ in the process of phase transition from coexistence ($P_{ext}$ tends to zero) to extinction ($P_{ext}$ approaches $1$). The critical value $p_c$ depends on the other parameters in the model, as well.

In Ref.~\cite{reichenbach}, the authors verified that the spiral wavelength increases with individual mobility and that the wavelength is proportional to $\sqrt{m}$. They found that there exists a critical mobility $M_c$. When mobility is greater than $M_c$, the pattern outgrows the system size, causing loss of biodiversity.

In this work, we obtain similar results in the case of fixed mobility and variable long-range connection probability $p$ or variable maximal interaction distance $R_{max}$. This means that increasing $p$ or $R_{max}$ is equivalent to increasing the mobility. Although the long-range interaction does not directly change the exchange rate $\epsilon$, it does change the spatial structure and leads to faster interactions, particularly for exchange. So, increasing $p$ or $R_{max}$ increases mobility $m$ indirectly. 

\begin{figure}
\includegraphics[width=0.4\textwidth]{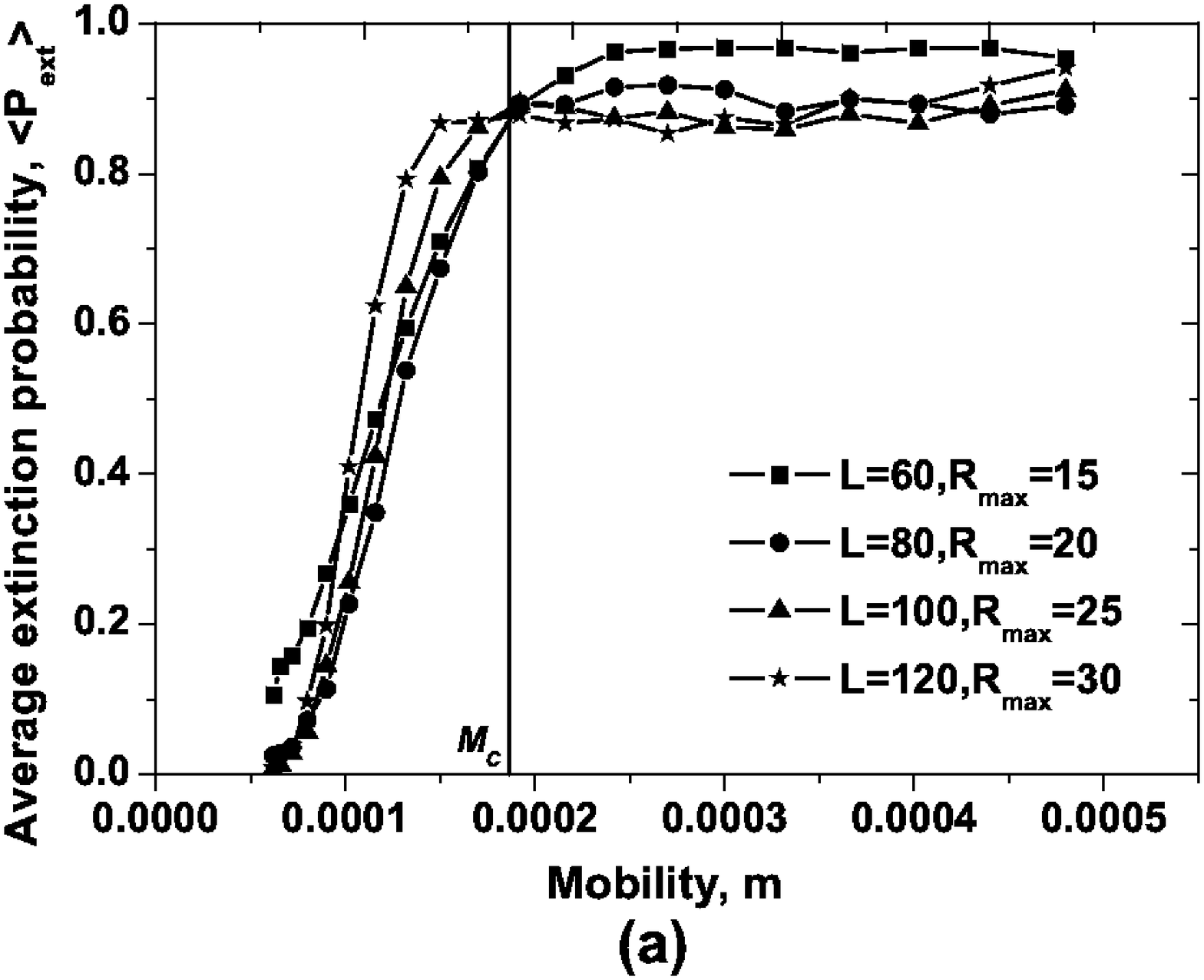}
\includegraphics[width=0.4\textwidth]{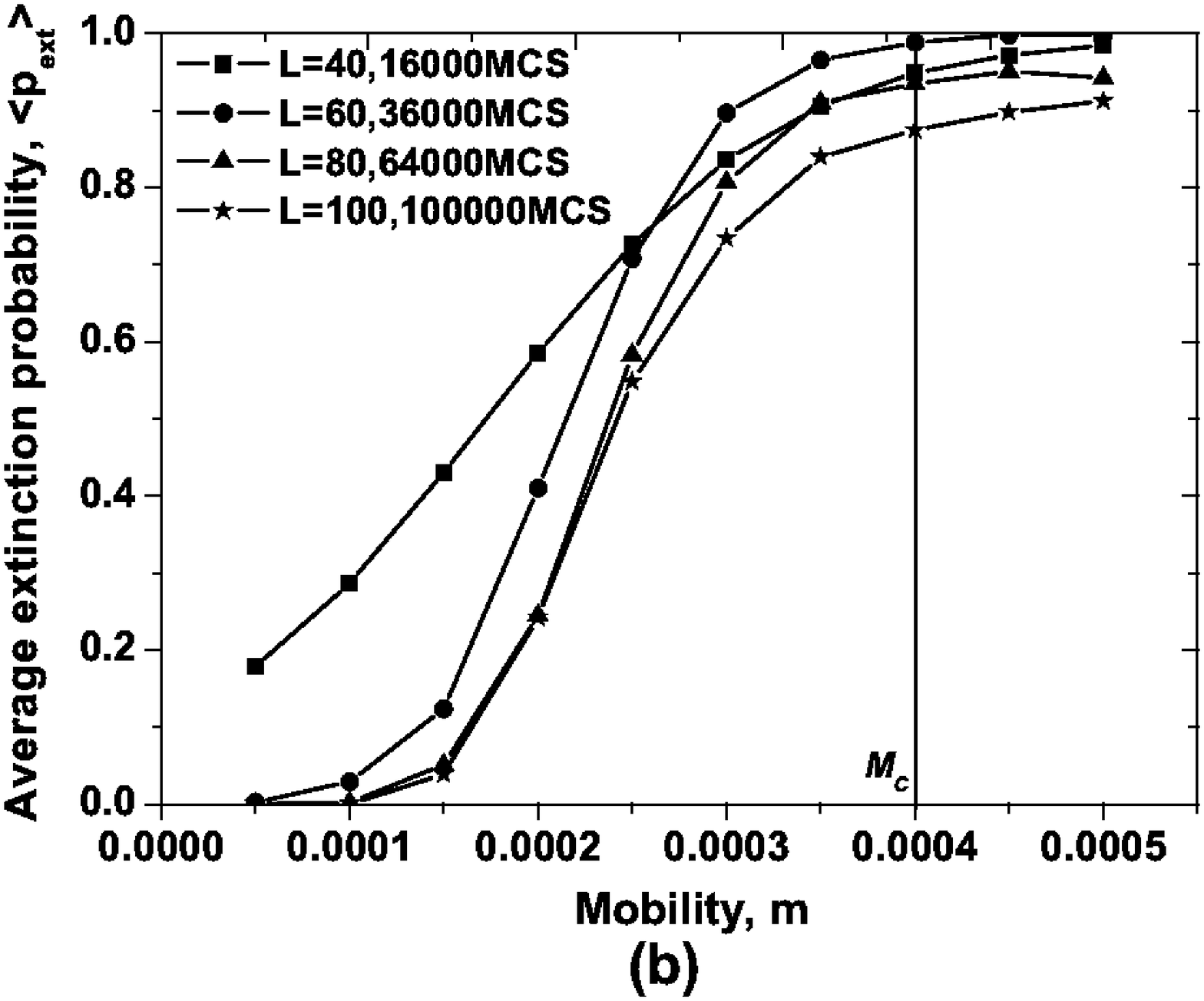}
\caption{The extinction probability as a function of mobility: (a) With long-range connections, the transition from stable coexistence to extinction sharpens at a critical mobility $M_c \approx 1.9 \times 10^{-4}$, $t = 10,000 MCS$. (b) Without long-range connections, $M_c \approx 4.0 \times 10^{-4}$. Parameters: $\mu = \sigma = 1$, $p = 0.02$, $R_{max} = L/4$.}
\label{fig06}
\end{figure}

In Fig.~\ref{fig06}(a), we plot the dependence of $P_{ext}$ on mobility $m$ in the presence of long-range interactions. With increasing mobility $m$, a sharpened transition emerges at a critical value $M_c \approx 1.9 \times 10^{-4}$, which is smaller than the value $(4.5 \pm 0.5) \times 10^{-4}$ provided in Ref.~\cite{reichenbach}. In Fig.~\ref{fig06}(b), we also compute $P_{ext}$ without long-range interactions ($p=0$), holding the other parameters the same in Fig.~\ref{fig06}(a). In these conditions, it takes much longer ($t = 10N$ MCS) to reach the critical value $M_c \approx 4 \times 10^{-4}$, which approximately coincides with the value $(4.5 \pm 0.5) \times 10^{-4}$. That is to say, in this case the system is more stable than with long-range interactions.

It is well known that long-range connections change spatial structure dramatically. To learn more information about the effect of long-range connections on the emerging spiral patterns, we computed the equal-time correlation function $g_{AA} \left( |r-r'| \right)$ at $r$ and $r'$ of species $A$ for the system's steady state, which is defined in Ref.~\cite{tobias}, as
\begin{equation}
g_{AA}\left( |r-r'| \right) = \left< a(r,t)a(r',t)) \right> - \left< a(r,t) \right> \left< a(r',t) \right>,
\label{eq05}
\end{equation}
where $\left< \ldots \right>$ stands for an average over all histories. 

\begin{figure}
\includegraphics[width=0.4\textwidth]{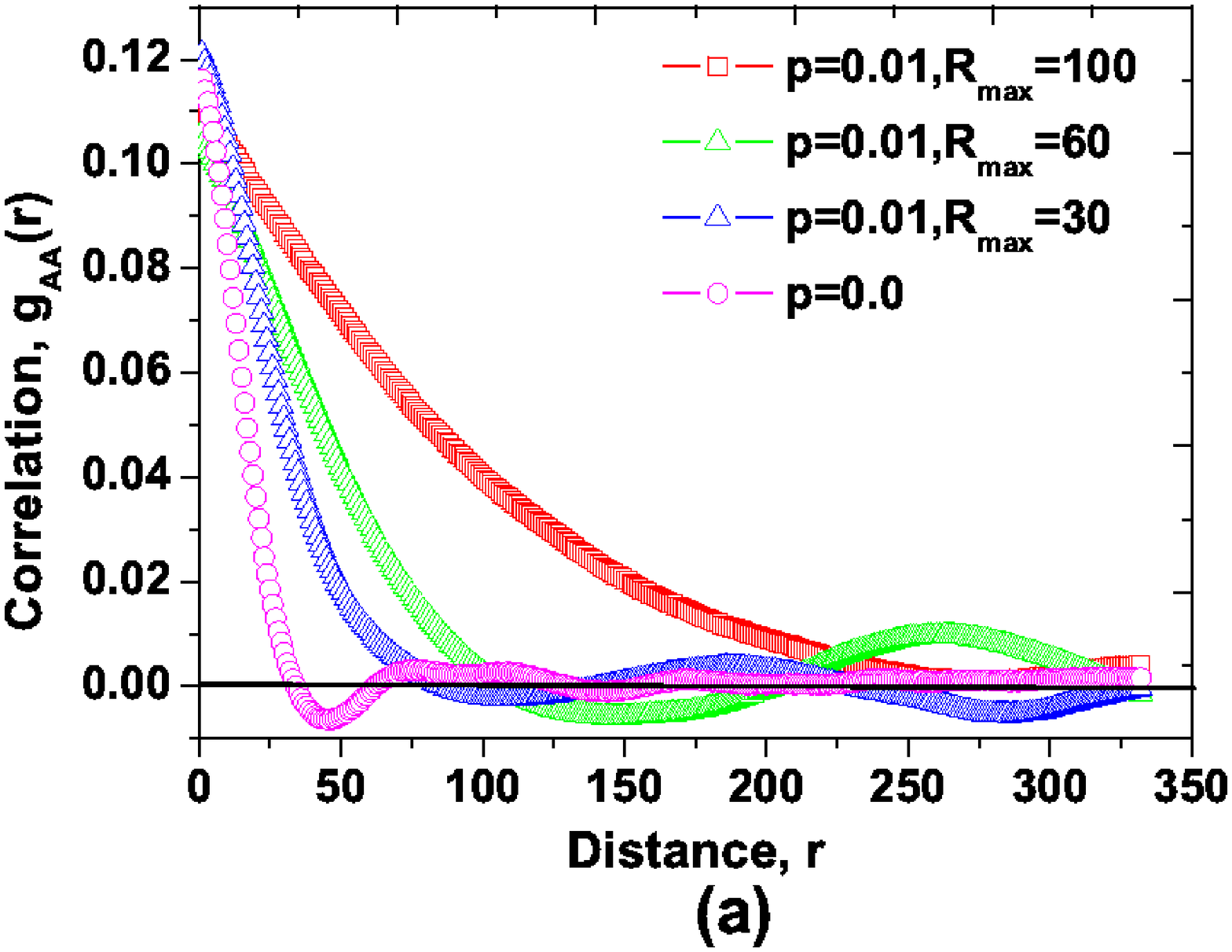}
\includegraphics[width=0.4\textwidth]{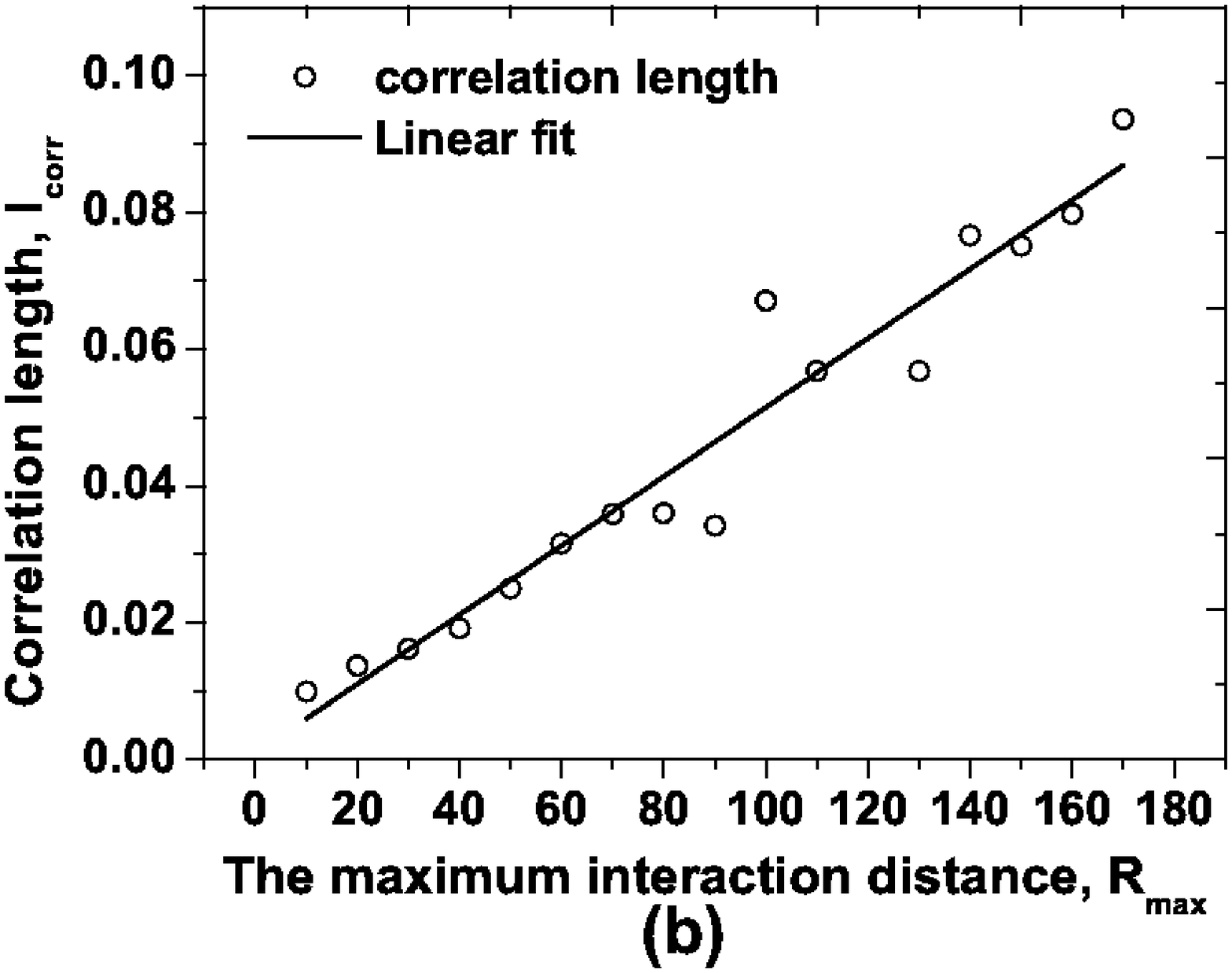}
\caption{(Color online) (a) The spatial correlation $g_{AA}(r)$ as a function of $r$ in the reactive steady state. (b) The dependence of correlation length $l_{corr}$ on the maximal long-range interaction distance $R_{max}$. Correlation length is depicted as circle. The black line is the linear fit. Parameters: $L = 1000$, $m = 1.2 \times 10^{-5}, t = 6000 MCS$, and $\mu = \sigma = 1$}
\label{fig07}
\end{figure}

Fig.~\ref{fig07}(a) plots $g_{AA}$ obtained from MC simulations. When the separating distance reaches zero, the correlation reaches its maximal value. With increasing distance, the correlation decreases and the spatial oscillations appear. This oscillation reflects the underlying spiraling spatial structures where the three species alternate in turn. Furthermore, the correlation functions could be characterized by their correlation length $l_{corr}$, which is the length at which the correlations decay by a factor $1/e$ from their maximal value. The value of $l_{corr}$ is obtained by fitting $g_{AA}(r)$ to exponentials $e^{-r/l_{corr}}$. This value conveys information on the typical size of the spirals~\cite{tobias2}. In Fig.~\ref{fig07}(b), we show the dependence of $l_{corr}$ on the maximal long-range interaction distance $R_{max}$. The results confirm the scaling relationship $l_{corr} \propto R_{max}$ for the fixed long-range connection probability $p$. In addition, it can be observed that the linear fit is less good when $R_{max}$ around 100. Through extra numerical simulations on the correlation length, we found that when $R_{max}$ is around or above 100, the system would be at extinction in a high probability. This could affect the correlation and correlation length.

In summary, we studied the influence of random long-range connection on four-state RPS games with NW networks based on extensive MC simulations. For a fixed maximal interaction distance $R_{max}$, as the probability of long-range connections $p$ increases, we observe that the spiral waves grow in size and (for larger $p$) disappear. When the spiral waves disappear, the system reaches a uniform state and biodiversity is lost. There exists a critical value $p_c$ separating coexistence from extinction. Similar behaviors are observed with increasing $R_{max}$ for a fixed $p$. To close more ecological realistic model, we also consider the case that $XE->EX$ occur with a smaller probability where $X$ and $E$ (empty place) are not neighboring sites. When $p$ or $R_{max}$ increases, we observe that both the size of spiral waves grow more slowly and the process of phase transition from coexistence to extinction are slower than before.

We compared the critical value $M_c$ obtained in two cases: with and without long-range connection. It is found that $M_c$ changes drastically and the systems becomes more unstable if even a weak long-range connection is presented. We conclude that long-range interactions could result in improved mobility, and it has dramatic effects on species coexistence. This point is also confirmed by the equal-time correlation functions for the system's steady state and by the correlation length for different $R_{max}$.

\bigskip
We are grateful to T. Reichenbach for helpful advice on simulation methods and thank referees' for their constructive suggestions. This work was supported by the National Natural Science Foundation of China under Grant No. $10305005$.

\end{document}